\documentclass[lettersize, journal]{IEEEtran}

\usepackage{mathrsfs}
\usepackage{amsmath}
\usepackage{amsfonts}
\usepackage{amsthm}
\usepackage{amssymb}
\usepackage{graphicx}
\usepackage{subfigure}
\usepackage{indentfirst}
\usepackage{array}
\usepackage{epstopdf}
\usepackage{cite}
\usepackage{mathrsfs}
\usepackage{enumerate}
\usepackage{bm}
\usepackage{dsfont}
\usepackage[linesnumbered,ruled,vlined]{algorithm2e}
\usepackage{pifont}
\usepackage{setspace}
\usepackage{multirow}
\usepackage{verbatim}
\usepackage{color}
\usepackage{stfloats}
\usepackage{makecell}
\usepackage{cancel}
\usepackage{multicol}
\usepackage{adjustbox}

\SetKwComment{Comment}{\color{black} $\triangleright$\ }{}
\SetKwInOut{KwIn}{Input}
\SetKwInOut{KwOut}{Output}
\SetKwProg{Proc}{Procedure}{}{}
\usepackage[colorlinks=true,linkcolor=blue,citecolor=blue]{hyperref}

\begin{document}
\bstctlcite{IEEEexample:BSTcontrol}

\title{\LARGE{NeRFCom: Feature Transform Coding Meets Neural Radiance Field for Free-View 3D Scene Semantic Transmission}}

\author{Weijie Yue, 
Zhongwei Si,
Bolin Wu,
Sixian Wang,
Xiaoqi Qin,
Kai Niu,
Jincheng Dai,
Ping Zhang

\thanks{This work was supported in part by the National Natural Science Foundation of China under Grant 62371063, Grant 62293481, Grant 62321001, Grant 92267301, Grant 61971062, in part by the Beijing Municipal Natural Science Foundation under Grant L232047, and Grant 4222012, in part by the National Key Research and Development Program of China under Grant 2024YFF0509700, in part by Program for Youth Innovative Research Team of BUPT No. 2023YQTD02, and in part by BUPT Excellent Ph.D. Students Foundation under Grant CX2022153. \emph{(Corresponding Authors: Jincheng Dai, Zhongwei Si.)}}

\thanks{Weijie Yue, Zhongwei Si, Bolin Wu, Sixian Wang, Kai Niu, and Jincheng Dai are with the Key Laboratory of Universal Wireless Communications, Ministry of Education, Beijing University of Posts and Telecommunications, Beijing 100876, China. (email: daijincheng@bupt.edu.cn, sizhongwei@bupt.edu.cn)}

\thanks{Xiaoqi Qin and Ping Zhang are with the State Key Laboratory of Networking and Switching Technology, Beijing University of Posts and Telecommunications, Beijing 100876, China.}

\thanks{Source code and model are available at \url{https://github.com/semcomm}.}

\vspace{-0.1em}
}

\maketitle

\begin{abstract}

We introduce NeRFCom, a novel communication system designed for end-to-end 3D scene transmission. 
Compared to traditional systems relying on handcrafted NeRF semantic feature decomposition for compression and well-adaptive channel coding for transmission error correction, our NeRFCom employs a nonlinear transform and learned probabilistic models, enabling flexible variable-rate joint source-channel coding and efficient bandwidth allocation aligned with the NeRF semantic feature’s different contribution to the 3D scene synthesis fidelity.
Experimental results demonstrate that NeRFCom achieves free-view 3D scene efficient transmission while maintaining robustness under adverse channel conditions.
\end{abstract}

\begin{IEEEkeywords}
Neural radiance field (NeRF), 3D scene transmission, semantic features, nonlinear transform coding.
\end{IEEEkeywords}

\section{Introduction}

\IEEEPARstart{V}{irtual} reality (VR) and augmented reality (AR) construct 3D scenes to provide users with immersive experiences \cite{latva2020key}.
However, traditional 3D scene synthesis techniques often rely on manual scene modeling, and the complex workflow increases the cost of deploying 3D technologies. Additionally, these methods require the transmission of large data volumes, typically ranging from hundreds of megabytes (MB) to gigabytes (GB), which places considerable burden on transmission systems.
Neural Radiance Field (NeRF) \cite{nerf} is an emerging 3D synthesis technique driven by machine learning that significantly reduces the need for manual operations. NeRF represents 3D scenes as continuous and differentiable functions, enabling the generation of scene images from arbitrary viewpoints through simple rendering.
Although NeRF can reduce the data volumes compared to traditional techniques, it demands tens to hundreds of MB data for 3D features. Moreover, the features exhibit complex correlations that traditional source coding schemes struggle to achieve efficient compression.
This highlights the need for advanced coding and transmission strategies for 3D scenes to further drive the development of VR and AR.

Current research on NeRF-based 3D scene transmission schemes are categorized into two paradigms: \emph{Post-processing with NeRF} and \emph{Integration with NeRF}.
The first paradigm follows a straightforward idea, leveraging existing image transmission techniques to send all observed images, which are then reconstructed at the receiver. NeRF is then applied as a post-processing step to synthesize the 3D scene. 
However, this approach leads to redundant data transmission as it overlooks the spatial correlations between images. Consequently, methods following the second paradigm are generally preferred, where the transmitter extracts NeRF 3D semantic features from sparse-view images, decomposes these 3D features into 2D feature planes, and enables free-view scene synthesis at the receiver \cite{DVGO, TensoRF, K-Planes}. Recent works \cite{VQNeRF, rho2023masked} further employ vector quantization and linear transforms on each 2D plane separately for feature compression. Nevertheless, their reconstruction performance depends on the vector indices-map coding, which suffers from ineffective redundancy reduction between indices, leading to low compression efficiency.
Meanwhile, these works focus solely on source compression, without consideration of transmission errors over wireless channels. Although they can be concatenated with advanced channel coding to mitigate the impact of error-prone channels, like \cite{wu2024semantic} introduced a framework for 3D human face transmission, the separate design of source and channel coding still leads to a ``cliff effect'', causing severe end-to-end performance degradation when the channel coding rate adjustment falls behind the channel.
In \cite{bourtsoulatze2019deep,deniz2022beyond}, a type of learned joint source and channel coding (JSCC) method was introduced to mitigate the cliff effect. While the effectiveness of JSCC methods have been validated for image and video transmission \cite{dai2022nonlinear, wang2023improved, 10158528, lyu2024semantic}, the high dimensionality and complex correlations of NeRF 3D features still present challenges for efficient and robust transmission.

To address the aforementioned challenges, we propose NeRFCom, a novel end-to-end communication framework for free-view 3D scene transmission.
Specifically, NeRFCom develops a neural-based nonlinear transform to encode 3D features into low-dimensional latent representations. A learned entropy model is then introduced to characterize the probability distribution of the latent representations. This model evaluates the information entropy, which reflects the contribution of each latent representation element to the overall fidelity of 3D scene synthesis.
Following this, NeRFCom incorporates a variable-rate neural JSCC module aligned with the entropy distribution to directly maps the latent representations into continuous-valued channel-input symbols, allowing for graceful performance degradation.
The entire system is optimized with a transmission rate-distortion (RD) objective, driven by neural networks implementing all NeRF and JSCC components, thereby facilitating end-to-end optimization.

\section{System Model and Methods}
\label{section_proposed_method}

\begin{figure*}[t]
\centering
\includegraphics[width=0.98\textwidth]{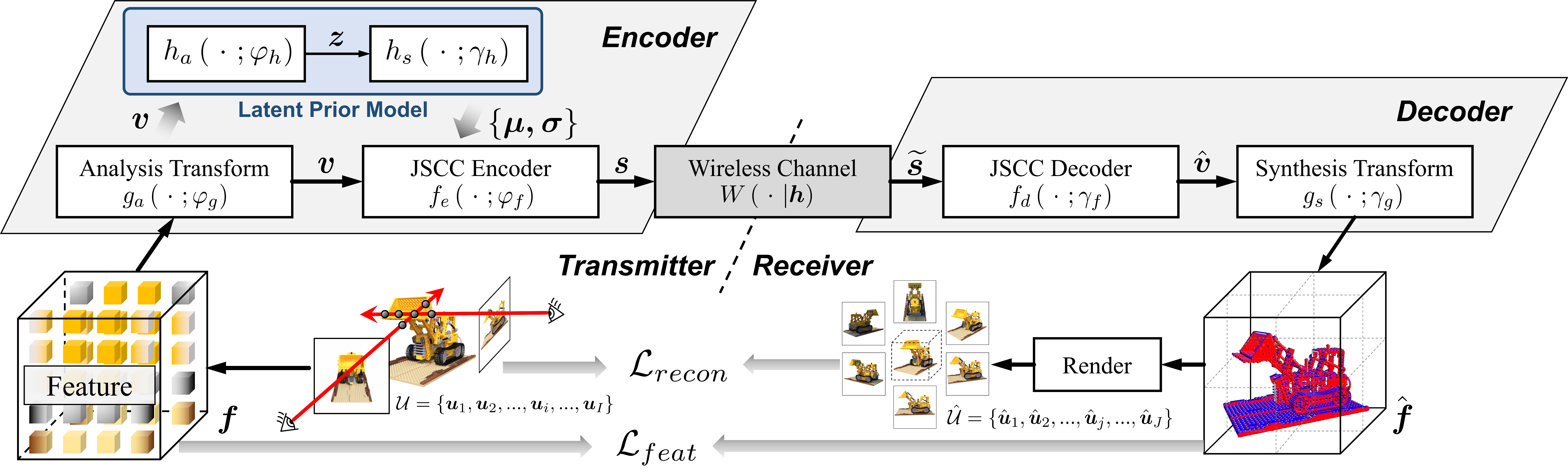}
\caption{The framework of our end-to-end free-view 3D scene transmission system.}
\label{fig_system_model}
\end{figure*}

\subsection{System Model}
The framework of the 3D scene transmission system is shown in Fig. \ref{fig_system_model}.
First, we extract a voxel-based 3D feature $\boldsymbol{f} \in \mathbb{R}^m$ from a set of sparse-view images $\mathcal{U}=\left\{ \boldsymbol{u}_1, \boldsymbol{u}_2, \ldots, \boldsymbol{u}_i, \ldots, \boldsymbol{u}_I \right\}$, where each $\boldsymbol{u}_i$ corresponds to an image observed from a specific viewpoint of 3D scene, and $I$ represents the total number of sparse-view images used for training. 
Next, $\boldsymbol{f}$ is fed into the Encoder to acquire variable-length channel symbols $\boldsymbol{s} \in \mathbb{R}^n$. Here, $m$ represents the dimensionality of $\boldsymbol{f}$, and $n$ corresponds to the number of channel symbols. The channel bandwidth ratio (CBR) is defined as $\text{CBR} = n / m$, which represents the ratio of transmitted channel symbols relative to the original feature dimension.
The details of the Encoder and Decoder are presented in Sec.~\ref{sec_encoder_decoder}.

Then the $\boldsymbol{s}$ is transmitted over the wireless channel $W\left( \cdot | \boldsymbol{h}\right) $, where $\boldsymbol{h}$ denotes the channel parameters. The received symbol is modeled as $\widetilde{\boldsymbol{s}} = W\left( \boldsymbol{s} | \boldsymbol{h} \right) = \boldsymbol{h}\boldsymbol{s} + \boldsymbol{n}$. The additive white Gaussian noise (AWGN) channel is considered in this paper, where each component of the noise vector $\boldsymbol{n}$ is independently sampled from a Gaussian distribution, $\boldsymbol{n} \sim \mathcal{N} \left( 0, \sigma^2 \mathbf{I} \right)$, and $\sigma^2$ is the average noise power.
Finally, the receiver decodes the features $\hat{\boldsymbol{f}} = \text{Decoder}(\widetilde{\boldsymbol{s}})$ from received symbols and utilizes these features to render $\hat{\mathcal{U}}=\left\{ \hat{\boldsymbol{u}}_1, \hat{\boldsymbol{u}}_2, \ldots, \hat{\boldsymbol{u}}_j, \ldots, \hat{\boldsymbol{u}}_J \right\}$, where $\hat{\boldsymbol{u}}_j$ corresponds to scene image rendered from a specific viewpoint, and $J$ represents the total number of synthesis images. Notably, $J\gg I$, as the system can synthesize images from viewpoints not included in the original set $\mathcal{U}$. 

The entire system is formulated as an optimization problem, where the synthesis images distortion is measured by the reconstruction loss $\mathcal{L}_\mathrm{recon}$. Additionally, a feature loss term $\mathcal{L}_\mathrm{feat}$ is imposed to measure feature distortion and regulate the bandwidth usage. This feature loss also serves as a regularization term, promoting stable training and facilitating faster convergence. The details are introduced in Sec.\ \ref{sec_opti}.

\subsection{Variable-Rate Feature Coded Transmission} \label{sec_encoder_decoder}
This section introduces the proposed variable-rate feature-coded transmission approach and provides details of the Encoder and Decoder as illustrated in Fig.\ \ref{fig_system_model}. The 3D voxel feature $\boldsymbol{f} \in \mathbb{R}^{D \times H \times W \times C}$, where $D$, $H$, and $W$ represent the number of voxels in depth, height, and width, respectively, and $C$ represents the number of feature channels per voxel.

The encoding process applies an analysis transform $g_a$ that maps $\boldsymbol{f}$ to a latent representation $\boldsymbol{v}$. The $g_a$ consists of 3D convolutional layers for progressive downsampling and reduction in the feature channel dimension, followed by patch merging to acquire transmission patches. The resulting representation is expressed as $\boldsymbol{v} = \{v_1, v_2, \dots, v_i, \dots, v_P\}$, where each $v_i$ is a transmission patch, and $P$ is the total number of patches.
Then $\boldsymbol{v}$ is fully factorized to derive the latent prior $\boldsymbol{z}$ using a lightweight neural network $h_a$. The latent prior $\boldsymbol{z}$ captures the critical distribution of $\boldsymbol{v}$ and guides the bandwidth allocation in the JSCC encoder. More resources are allocated to regions that contribute significantly to 3D scene synthesis.
Finally, the encoded channel symbols are represented as $\boldsymbol{s}=\left\{s_1, s_2, \cdots, s_i,\cdots,s_P\right\}$, where each $s_i$ is a variable-length symbol corresponding to a transmission patch $v_i$.

Specifically, the entropy model $p_{\boldsymbol{v}|\boldsymbol{z}}(\boldsymbol{v}|\boldsymbol{z})$ describes the importance diversity of $\boldsymbol{v}$ and is formulated as:
\begin{equation}
    \begin{aligned}
    p_{\boldsymbol{v}|\boldsymbol{z}}(\boldsymbol{v}|\boldsymbol{z}) = \prod_{i} \left(\mathcal{N}(v_{i}|\boldsymbol{z}) * \mathcal{U}(-\frac{1}{2}, \frac{1}{2})\right)(v_{i}),
    \end{aligned}
\end{equation}
where the convolutional operation ``$* $’’ with a standard uniform distribution is used to align the prior with the marginal distribution. The latent prior $\boldsymbol{z}$ provides prior information for estimating the probability distribution of $\boldsymbol{v}$, which is essential for efficient entropy coding and bandwidth allocation. The entropy of each patch $v_i$, calculated as $-\log p_{v_i|\boldsymbol{z}}(v_i|\boldsymbol{z})$, is used to allocate bandwidth resources $\bar{k}_{i}$ according to:
\begin{equation}
	\bar{k}_{i}= Q\Big(-\eta\operatorname{log}p_{v_{i}|\boldsymbol{z}}(v_{i}|\boldsymbol{z})\Big),
\end{equation}
where $Q(\cdot)$ is a scalar quantization function, and $\eta$ is a scaling factor. The range of quantization and the value of $\eta$ can be adjusted to control the bandwidth resource used. This ensures that more bandwidth is allocated to patches with higher entropy, which are more critical for 3D scene synthesis.

A dynamic neural network \cite{han2021dynamic} is employed in both the JSCC encoder $f_e$ and decoder $f_d$ to realize variable-length coding. The output layer dimension is determined by $\bar{k}_{i}$, where more important features are corresponding to higher-dimensional outputs.
At the receiver, a series of inverse operations are applied to decode and reconstruct the transmitted features. The JSCC decoder $f_d$ reconstructs the latent representation $\boldsymbol{\hat{v}}$ from the received symbols $\boldsymbol{\widetilde{s}}$. The synthesis transform $g_s$ then performs the inverse operation to recover the 3D feature $\boldsymbol{\hat{f}}$, enabling the receiver to render images from any desired viewpoint.

To summarize, the entire symbol generation and recovery process is as follows:
\begin{align*}
\boldsymbol{f}\xrightarrow{g_{a}(\cdot)}\boldsymbol{v}\xrightarrow{f_{e}(\cdot)}\boldsymbol{s}\xrightarrow{W(\cdot|\boldsymbol{h})}\boldsymbol{\tilde{s}}\xrightarrow{f_{d}(\cdot)}\boldsymbol{\hat{v}}\xrightarrow{g_{s}(\cdot)}\boldsymbol{\hat{f}}, \\
\text{with the latent prior } \boldsymbol{v}\xrightarrow{h_{a}(\cdot)}\boldsymbol{z}\xrightarrow{h_{s}(\cdot)}\{\boldsymbol{\mu},\boldsymbol{\sigma}\}.
\end{align*}

Here, $g_a(\cdot)$ extracts latent features $\boldsymbol{v}$ from the input 3D feature $\boldsymbol{f}$, $f_e(\cdot)$ encodes $\boldsymbol{v}$ into channel symbols $\boldsymbol{s}$, $W(\cdot|\boldsymbol{h})$ represents the noisy wireless channel, $f_d(\cdot)$ decodes the received symbols $\boldsymbol{\tilde{s}}$ into $\boldsymbol{\hat{v}}$, and $g_s(\cdot)$ reconstructs the final 3D feature $\boldsymbol{\hat{f}}$. Additionally, the latent prior modeling components $h_a(\cdot)$ and $h_s(\cdot)$ estimate the distribution of latent features, where $h_a(\cdot)$ encodes $\boldsymbol{v}$ into $\boldsymbol{z}$, and $h_s(\cdot)$ analyzes $\boldsymbol{z}$ to acquire the distribution parameters $\{\boldsymbol{\mu}, \boldsymbol{\sigma}\}$.

\begin{algorithm}[t]\label{alg_training}
\caption{Multi-stage Training of NeRFCom}
\KwIn {Spare view images $\mathcal{U}=\left\{ \boldsymbol{u}_1,\boldsymbol{u}_2,\cdots,\boldsymbol{u}_I \right\}$, \\
viewpoint direction $\boldsymbol{d}$}
\KwOut {Render image $\hat{\boldsymbol{u}}$ of 3D scene for given $\boldsymbol{d}$}

\Proc{\rm \textbf{{\hspace{-0.118em}1:}} {Update NeRF}}{
\For(\Comment*[f]{Iteration}){$t = 1, 2, \dots, T_1$ }
{
    $\boldsymbol{f} \gets \boldsymbol{d}$ \Comment*[f]{Acquire 3D feature}

    $\hat{\boldsymbol{u}} \gets \text{Render}(\boldsymbol{f})$

    $\mathcal{L}_\mathrm{recon} \gets d_{\text{MSE}}(\boldsymbol{u}, \hat{\boldsymbol{u}})$

    Update the NeRF model
}
\Return Updated NeRF model
}

\Proc{\rm \textbf{{\hspace{-0.118em}2:}} {Fix NeRF, update Encoder and Decoder}}{
\For(\Comment*[f]{Iteration}){$t = 1, 2, \dots, T_2$ }
{
    $\boldsymbol{f} \gets \text{Updated NeRF model}$

    $\boldsymbol{s} \xleftarrow{f_e} \boldsymbol{v} \xleftarrow{g_a} \boldsymbol{f}$
    \Comment*[f]{Map 3D feature to channel symbol}

    $\widetilde{\boldsymbol{s}} \xleftarrow{W} \boldsymbol{s}$
    \Comment*[f]{Symbol Transmission}
    
    $\hat{\boldsymbol{f}} \xleftarrow{g_s} \hat{\boldsymbol{v}} \xleftarrow{f_d} \widetilde{\boldsymbol{s}}$
    \Comment*[f]{Map channel symbol to 3D feature}

    $\mathcal{L}_\mathrm{feat} \gets d_{\text{MSE}}(\boldsymbol{f}, \hat{\boldsymbol{f}}) + \lambda (R_{\boldsymbol{v}} + R_{\boldsymbol{z}})$
    
    Update the Encoder and Decoder model
}
\Return Updated Encoder and Decoder model
}

\Proc{\rm \textbf{{\hspace{-0.118em}3:}} {Finetune NeRF, fix Encoder and Decoder}}{
\For(\Comment*[f]{Iteration}){$t = 1, 2, \dots, T_3$ }
{
    $\boldsymbol{f} \gets \boldsymbol{d}$
    \Comment*[f]{Acquire 3D feature}

    $\boldsymbol{s} \gets \text{Encoder}(\boldsymbol{f})$

    $\widetilde{\boldsymbol{s}} \xleftarrow{W} \boldsymbol{s}$
    \Comment*[f]{Symbol Transmission}
    
    $\hat{\boldsymbol{f}} \gets \text{Decoder}(\widetilde{\boldsymbol{s}})$

    $\hat{\boldsymbol{u}} \gets \text{Render}(\hat{\boldsymbol{f}})$

    $\mathcal{L}_\mathrm{recon} \gets d_{\text{MSE}}(\boldsymbol{u}, \hat{\boldsymbol{u}})$

    Update the NeRF model
}
\Return Render image $\hat{\boldsymbol{u}}$ of 3D scene for given $\boldsymbol{d}$
}
\end{algorithm}

\subsection{Optimization Goal and Strategy}
\label{sec_opti}
The optimization of the entire system encompasses both the 3D features transmission and scene synthesis result.
The former focuses on optimizing transmission loss under entropy constraints, while the latter aims to enhance the end-to-end scene synthesis fidelity.

\subsubsection{Feature Transmission Loss}
The optimization of feature transmission creates a variational density $q_{\boldsymbol{s,z|f}}$ to approximate the true posterior $p_{\boldsymbol{s,z|f}}$ \cite{balle2017end, balle2018variational}. We use the Kullback-Leibler (KL) divergence to quantify the discrepancy between two distributions. The feature loss can be formulated as
\begin{equation}
    \begin{aligned}
    \mathcal{L}_\mathrm{feat}&=\mathbb{E}_{\boldsymbol{f}\sim p_{\boldsymbol{f}}}D_{\mathrm{KL}}(q_{\boldsymbol{s},\boldsymbol{z}|\boldsymbol{f}}\|p_{\boldsymbol{s},\boldsymbol{z}|\boldsymbol{f}}) \\
    &=\lambda\left(-\log p_{\boldsymbol{v}|\boldsymbol{z}}(\boldsymbol{v}|\boldsymbol{z})-\log p_{\boldsymbol{z}}({\boldsymbol{z}})\right)+d_{\text{MSE}}(\boldsymbol{f},{\hat{\boldsymbol{f}}}) \\
    &=\lambda(R_{\boldsymbol{v}}+R_{\boldsymbol{z}})+d_{\text{MSE}}(\boldsymbol{f},{\hat{\boldsymbol{f}}}),
    \end{aligned}
    \label{equ_feat_loss}
\end{equation}
where the first two terms represent the channel bandwidth costs, $R_{\boldsymbol{v}}$ and $R_{\boldsymbol{z}}$, for the latent representation $\boldsymbol{v}$ and side information $\boldsymbol{z}$, respectively. Specifically, the term $-\log p_{\boldsymbol{v}|\boldsymbol{z}}(\boldsymbol{v}|\boldsymbol{z})$ captures the entropy cost associated with encoding the latent representation $\boldsymbol{v}$, conditioned on the latent prior $\boldsymbol{z}$. Similarly, $-\log p_{\boldsymbol{z}}(\boldsymbol{z})$ reflects the entropy cost of encoding the latent prior $\boldsymbol{z}$ itself. The second term in the loss function quantifies the distortion in feature transmission, where the distortion function $d_{\text{MSE}}(\boldsymbol{f}, \hat{\boldsymbol{f}})$ is defined as the mean squared error (MSE) between the original feature $\boldsymbol{f}$ and the reconstructed feature $\hat{\boldsymbol{f}}$. Finally, the multiplier $\lambda$ controls the trade-off between the total channel bandwidth cost and the feature distortion.

\subsubsection{End-to-End Reconstruction Loss}
NeRF-based methods aim to optimize a continuous scene function from a set of images to achieve 3D reconstruction. This scene function is utilized to render the final image $\boldsymbol{\hat{u}}$ from a specific viewpoint.
To assess the accuracy of the rendered result, it is compared to the ground truth image $\boldsymbol{u}$. By considering pixel alignment between the two images, we quantify the reconstruction loss $\mathcal{L}_\mathrm{recon}$ using the mean squared error (MSE), which is defined as:
\begin{equation}
\mathcal{L}_\mathrm{recon}=d_{\text{MSE}}\left( \boldsymbol{u}, \hat{\boldsymbol{u}} \right).
\end{equation}

The overall system can be formulated as a rate-distortion (RD) optimization problem, with the objective function
\begin{equation}
\begin{aligned}
	\mathcal{L}&=\mathcal{L}_\mathrm{feat}+\mathcal{L}_\mathrm{recon} \\
	&=\lambda \left( R_{\boldsymbol{v}}+R_{\boldsymbol{z}} \right) 
	+d_{\text{MSE}}( \boldsymbol{f},\hat{\boldsymbol{f}} )
	+d_{\text{MSE}}\left( \boldsymbol{u}, \hat{\boldsymbol{u}} \right),
\end{aligned}
\label{equ_loss}
\end{equation}
where $R_{\boldsymbol{v}}$ and $R_{\boldsymbol{z}}$ correspond to the bandwidth terms specified in Eq. \eqref{equ_feat_loss}, and we employ the MSE criterion to ensure the quality of both feature transmission and final reconstruction.

\begin{figure*}[t]
\centering
\includegraphics[width=\textwidth]{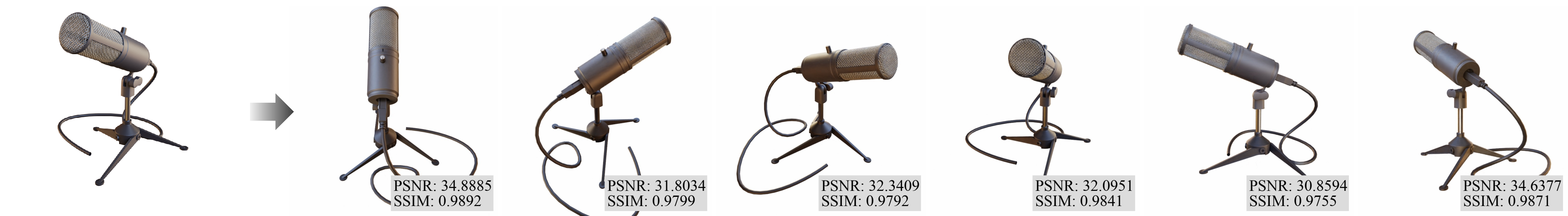}
\caption{Free-view 3D scene transmission results. The first image depicts the original 3D scene (a microphone), and the subsequent images display the rendering results from various viewpoints using the transmitted features.}
\label{fig_res_nerfcom}
\end{figure*}

\section{Experimental Results}\label{section_experiments}
In this section, we first introduce the experimental setup and then present our experimental results.
\subsection{Experimental Setup}

\subsubsection{Training Details}
Our model is trained on the Synthetic-NeRF dataset \cite{nerf}, which contains eight 3D scenes, each with 100 training images and 200 testing images at a resolution of $800\times800$. We adopt a progressive training strategy comprising three phases. First, 3D features are acquired using established techniques such as \cite{nerf, DVGO, TensoRF, K-Planes}. Next, we train the Encoder and Decoder components, which are central to achieving efficient compression and adaptive transmission. Finally, the Encoder and Decoder are fixed, and we fine-tune the 3D reconstruction model to enhance output quality. The complete step-by-step procedure is provided in Algorithm \ref{alg_training}.

The entire system is trained for 90,000 iterations, with procedure transitions happening at iterations 30,000 and 60,000. The Adam optimizer is used, along with an exponential learning rate decay schedule that includes a warm-up phase. All experiments are conducted on a single NVIDIA 3090 Ti GPU.

\subsubsection{Comparison Schemes}
We compare our system with DVGO \cite{DVGO}, TensoRF \cite{TensoRF}, and K-Planes \cite{K-Planes}, as well as schemes that incorporate classic data compression techniques. Specifically, \cite{VQNeRF, rho2023masked} employ vector quantization (VQ) and linear transformation coding (LTC) to compress features, which we refer to as ``VQ-NeRF'' and ``LTC-NeRF'' in discussions.
Following the separation designs, we combine these techniques with adaptive modulation and coding (AMC) mechanism. The channel coding rate and modulation order are selected based on the modulation and coding scheme (MCS) Table from the 3GPP TS 38.214 standard. For simplicity, we use ``+'' to denote these combinations, for instance, ``DVGO + 5G LDPC'' refers to DVGO combined with 5G LDPC channel coding. Unless otherwise noted, the LDPC code length is set to 4096.

\subsubsection{Metrics}
We use the widely adopted image metrics peak signal-to-noise ratio (PSNR) and structural similarity (SSIM) index to evaluate the distortion performance; higher values indicate better performance. For bitrate metrics, we use the CBR, which reflects the efficiency of bandwidth utilization in our transmission system. A higher CBR indicates greater bandwidth resource requirements, while a lower CBR signifies more efficient compression and transmission.

\subsection{Experimental Results}

\begin{figure}[t]
\centering
\includegraphics[width=0.98\columnwidth]{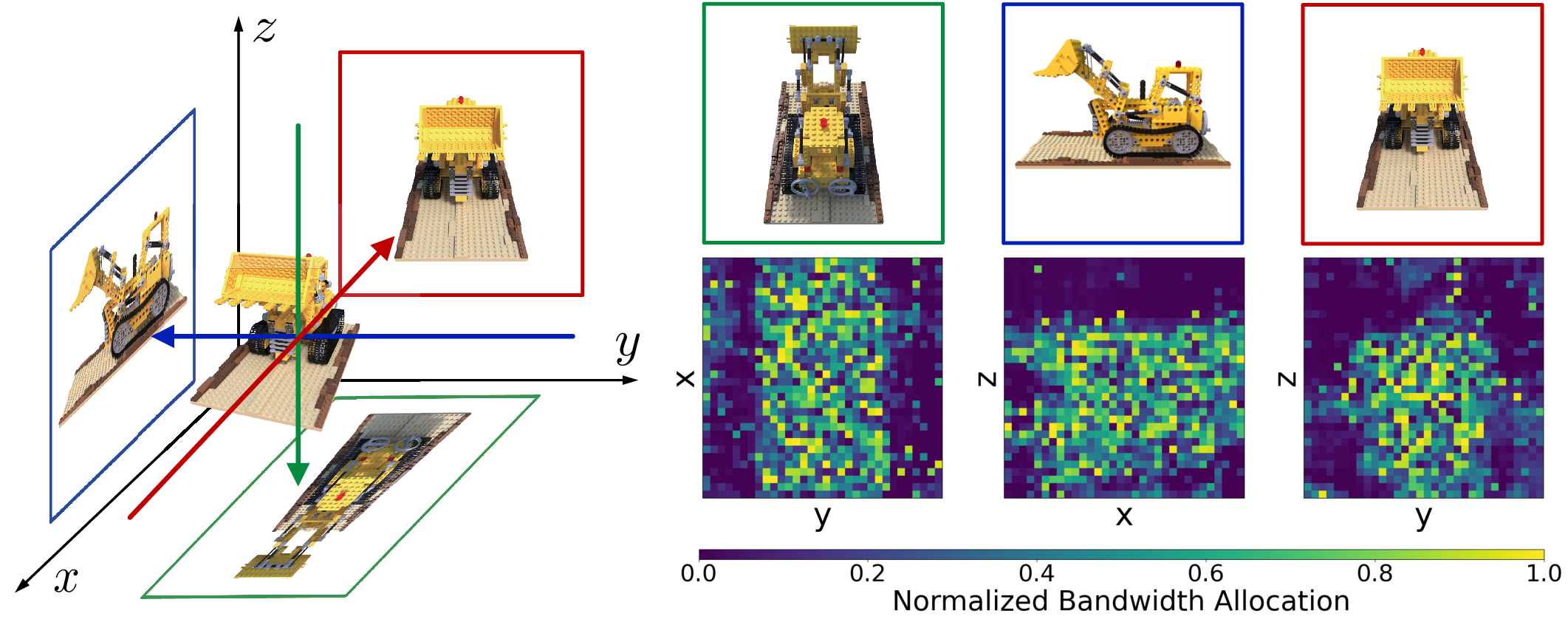}
\caption{Bandwidth allocation for different view images. Less bandwidth is allocated to blank areas in the viewpoint images, while more bandwidth is allocated to scene-relevant areas. The allocation is correlated with the spatial distribution of NeRF features.
}
\label{fig_res_heatmap}
\end{figure}

\begin{figure}[t]
\centering
\includegraphics[width=0.98\columnwidth]{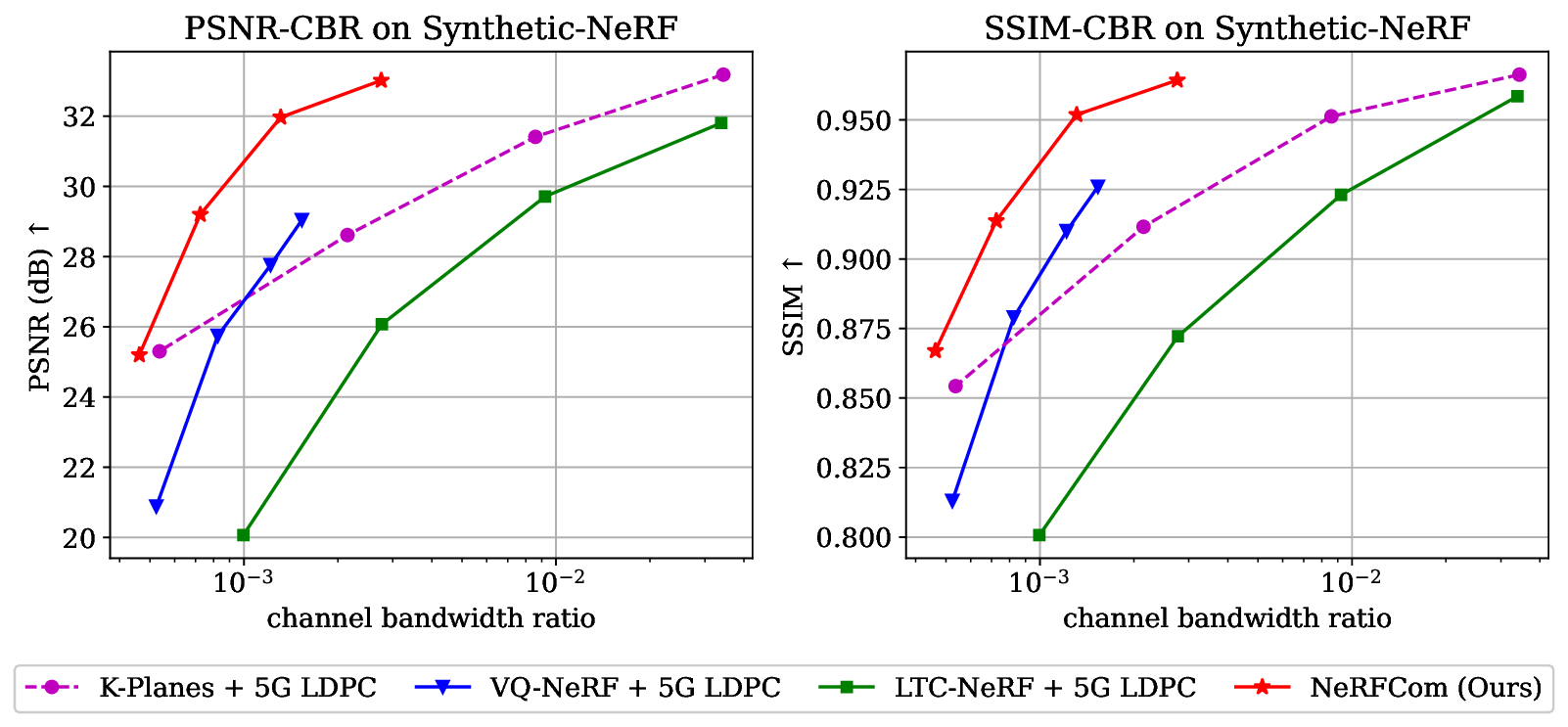}
\caption{PSNR and SSIM performance versus the average channel bandwidth ratio (CBR) at channel $\text{SNR}=10 \text{ dB}$.}
\label{fig_res_rd}
\end{figure}

\begin{figure*}[t]
	\centering
	\includegraphics[width=0.96\textwidth]{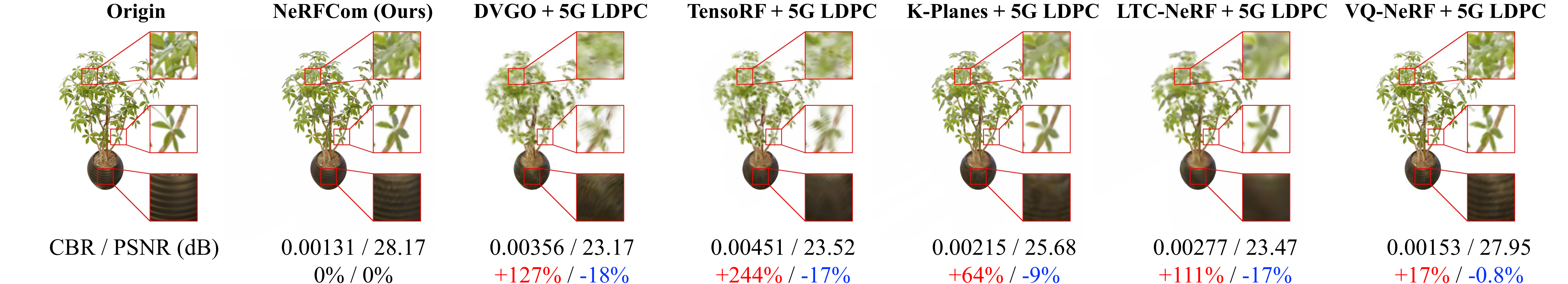}
	\caption{Examples of visual comparisons, with each result zoomed in on three specific details. The corresponding metric values and relative percentages are displayed below each result.}
	\label{fig_res_vis_comparisons}
\end{figure*}

\begin{figure}[t]
\centering
\includegraphics[width=0.9\columnwidth]{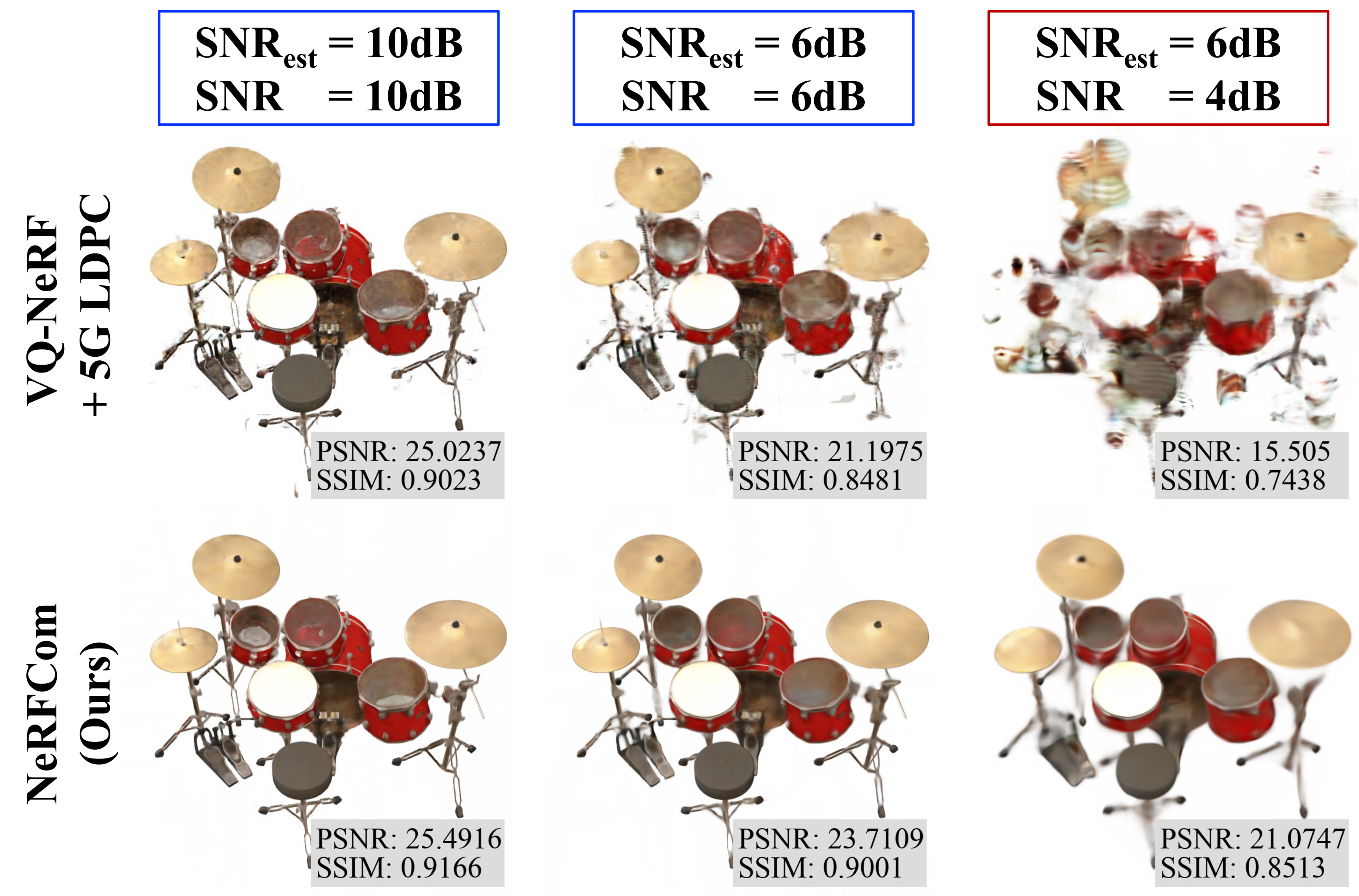}
\caption{A Example showing graceful degradation as channel SNR decreases.}
\label{fig_res_snrs}
\end{figure}

\subsubsection{Free-View 3D Scene Transmission Results}
We first demonstrate the free-view 3D scene transmission results of our NeRFCom system in Fig. \ref{fig_res_nerfcom}. The transmitter extracts the features of the 3D scene from sparse-view images, while the receiver renders the scene from any viewpoint.

\subsubsection{Variable-Rate Feature Coding Results}
The features exhibit spatial redundancy, with some parts being irrelevant for reconstruction but still consuming bandwidth.
Fig. \ref{fig_res_heatmap} presents the variable-rate feature coding results. On the left, the 3D scene is viewed from three directions (green, red, and blue rays in the coordinate system). The right side shows the corresponding projection images and bandwidth allocation. Yellow-green areas indicate higher resource allocation, while dark blue represents lower allocation. The results show that our system prioritizes bandwidth for scene-relevant regions, allocating minimal or no resources to empty areas, thereby improving transmission reliability and efficiency.

\subsubsection{Rate-Distortion Performance}
Fig.\ \ref{fig_res_rd} shows the RD performance at channel $\text{SNR}=10\text{dB}$. We can control bandwidth consumption by adjusting the dimensionality of the 3D features.
While “VQ-NeRF + 5G LDPC” outperforms other comparisons at high CBR level, this method degrade as the proportion of codebook overhead increases, our NeRFCom delivers superior RD performance.
Fig.\ \ref{fig_res_vis_comparisons} shows the visualized results of each scheme over the AWGN channel with $\text{SNR}=10 \text{dB}$. An AMC scheme using a 2/3 rate (4096, 6144) LDPC code with 16-QAM modulation is employed. For results, we zoom in three details and show the metric values below the images.
As seen from the results, our NeRFCom reconstructs more accurate details, such as clearer edges of the leaves.

We selected the highest-performing RD scheme ``VQ-NeRF + 5G LDPC'' for comparison to evaluate performance across various channel SNRs.Fig. \ref{fig_res_snrs} demonstrates the results across various SNR from $10 \text{dB}$ to $6 \text{dB}$ at a low $\text{CBR} = 0.0015$. 
The first two columns show cases where the estimated $\text{SNR}_\text{est}$ matches the actual channel $\text{SNR}$. In these cases, the ``VQ-NeRF + 5G LDPC'' results exhibit artifacts as the channel condition degrades, but the main structure of the 3D scene is still reconstructed. The last column presents a situation where the estimated $\text{SNR}_\text{est}$ does not match the actual channel $\text{SNR}$, causing ``VQ-NeRF + 5G LDPC'' to fail in reconstructing the 3D scene due to insufficient channel coding protection.
In contrast, NeRFCom experiences only a loss of fine details as channel conditions worsen, while maintaining superior overall reconstruction quality and exhibiting more graceful performance degradation.

\section{Conclusion}\label{section_conclusion}
In this paper, we have proposed NeRFCom, designed for 3D scene semantic transmission. NeRFCom takes sparse view images as input and allows the receiver to synthesize images from any viewpoint.
Our system incorporates variable-rate coding to realize efficient transmission of complex 3D features, which dynamically allocates more bandwidth to features that contribute more significantly to synthesis fidelity.
Furthermore, the integrated JSCC framework ensures graceful performance degradation under adverse channel conditions.
Experimental results demonstrate the efficiency and robustness of our 3D scene transmission system.
A limitation is the longer training time, and the need for per-scene optimization. However, inference remains efficient, allowing pre-trained models to be deployed. Future work will explore faster training and large-scale datasets to improve efficiency and generalization for more potential applications.

\bibliographystyle{IEEEtran}
\bibliography{CL_main.bib}

\vfill

\end{document}